**Polarization-diverse soliton transitions and deterministic switching dynamics in strongly-coupled and self-stabilized microresonator frequency combs**

Wenting Wang[1,2]*, Heng Zhou[3], Xinghe Jiang[1], Tristan Melton[1], Abhinav Kumar Vinod[1], Mingbin Yu[4,5], Guo-Qiang Lo[5], Dim-Lee Kwong[5], and Chee Wei Wong[1,*]

[1] Fang Lu Mesoscopic Optics and Quantum Electronics Laboratory, University of California, Los Angeles, CA 90095, United States of America

[2] Communication and Integrated Photonics Laboratory, Xiongan Institute of Innovation, Chinese Academy of Sciences, Xiong'an New Area 071700, China

[3] Key Lab of Optical Fiber Sensing and Communication Networks, University of Electronic Science and Technology of China, Chengdu 611731, China

[4] Shanghai Institute of Microsystem and Information Technology, Shanghai, China

[5] Institute of Microelectronics, A*STAR, Singapore 117865, Singapore

*Email: wenting.wang@xii.ac.cn; cheewei.wong@ucla.edu

Dissipative Kerr soliton microcombs in microresonators has enabled fundamental advances in chip-scale precision metrology, communication, spectroscopy, and parallel signal processing. Here we demonstrate polarization-diverse soliton transitions and deterministic switching dynamics of a self-stabilized microcomb in a strongly-coupled dispersion-managed microresonator driven with a single pump laser. The switching dynamics are induced by the differential thermorefractivity between coupled transverse-magnetic and transverse-electric supermodes during the forward-backward pump detunings. The achieved large soliton existence range and deterministic transitions benefit from the switching dynamics, leading to the cross-polarized soliton microcomb formation when driven in the transverse-magnetic supermode of the single resonator. Resultantly the pump laser always exists at the effective blue detuning of the transverse-magnetic resonance, fundamentally mitigating the thermal destabilization barrier and improving accessibility of the soliton formation regime. Subsequently and secondly, we demonstrate two distinct polarization-diverse soliton formation routes – arising from chaotic or periodically-modulated waveforms via pump power selection. The generated self-stabilized supermode microcomb features an extraordinarily large soliton existence range, a variety of soliton state transitions with well-defined pump laser tuning, high pump-microcomb conversion **efficiency, and low repetition rate phase noise. Thirdly, to observe the cross-polarized supermode transition dynamics, we develop a parametric temporal magnifier with picosecond resolution, MHz frame rate and sub-ns temporal windows. We construct picosecond temporal transition portraits in 100-ns recording length of the strongly-coupled solitons, mapping the transitions from multiple soliton molecular states to singlet solitons. This study underpins polarization-diverse soliton microcombs for chip-scale ultrashort pulse generation, supporting applications in frequency and precision metrology, communications, spectroscopy and information processing.**

Dissipative solitons are optical pulses generated in nonlinear cavity resonators that sustain their temporal and spectral shape through Kerr nonlinearity and wave dispersion as well as power dissipation and parametric gain. The generation of dissipative solitons in monolithic microresonators [1-5] has opened a new research area to realize chip-scale optical frequency combs and explore integrated spatiotemporal light localization and ultrafast dynamics. A variety of high-$Q$ microresonators with unprecedented compactness and complementary metal-oxide-semiconductor compatibility such as semiconductor-based resonators [6,7], crystalline resonators [2], and silicon wedge resonators [8] have been demonstrated to enable the formation of dissipative solitons. By externally driving millimeter- or micrometer-scale resonators with a continuous-wave (CW) [2] or synchronized pulse laser [9], dissipative solitons can spontaneously emerge from nonlinear stochastic or periodic waveform backgrounds and be sustained by the CW or pulse background excitation. Due to their octave-spanning spectral coverage, high phase coherence, and scalable repetition rates spanning from microwave to terahertz domains, dissipative soliton microcombs have attracted significant attention recently as coherent broadband light sources. They have been applied to abundant proof-of-principle system-level applications such as optical frequency metrology [10,11], massively-parallel multichannel communication [12-14], laser spectroscopy [15-17], precision distance metrology [18-20], low-noise terahertz wave and microwave generation [21-23], astronomical spectroscopy [24,25], parallel coherent laser ranging [26], and convolutional processing networks [27,28].

In addition to the wide-reaching applications, rich and complex soliton dynamics have been explored in integrated microresonators, establishing many soliton modalities. By controlling pump laser power and pump-resonance detuning for example, the dynamics can be elucidated with a two-dimensional stability map, depending on the solutions of the one-dimensional Lugiato-Lefever equation or its modifications, after considering perturbations such as avoided-mode crossings and the material ultrafast response. Typically the number of solitons formed in microresonators are inherently stochastic and the soliton dynamics cannot be manipulated deterministically [2]. By backward-tuning pump laser wavelength, continuous soliton switching dynamics from multiple soliton states to a single soliton state are revealed, related to power-dependent resonance frequency shift induced by thermo-optic effects [29]. Furthermore, a modulation of intracavity CW background physically arranges the intracavity nonlinear field to form temporally ordered ensembles of soliton crystals and molecules [30,31]. The corresponding optical spectra are carved with largely enhanced comb lines spaced by multiple free spectral ranges through avoided-mode crossing. Apart from shape-invariant temporal stationary solitons, non-stationary soliton states exhibiting a spatially periodic oscillatory behavior [32,33] can be excited by controlling the pump laser power, detuning or additional avoided-mode crossings [34]. Soliton trapping is observed in the microresonator [35] by optimizing intracavity field interaction in the spatial and temporal domains between the distinct transverse mode families. Recently, observations of dissipative solitons in coupled microresonators [36,37] offer new opportunities to examine emergent nonlinear soliton dynamics. The formation physical process and ultrafast dynamics of dissipative soliton in coupled microresonators remain however largely unexplored.

In the soliton formation process, sizable cavity thermorefractive transients occur at microsecond to millisecond time-scales in the microresonators due to heating from the intracavity pump CW laser – the key obstacle that limits the accessibility of the dissipative soliton microcomb, especially the single-soliton microcomb. The generation of single soliton microcombs remains a major technical challenge due to the presence of strong thermally induced resonance shifts. Rapidly sweeping the pump laser frequency determined by the thermal time constant enables quick transition through the modulation instability regime to avoid the thermal transients [2]. Two-step power kicking has also been applied to suppress detrimental laser heating to keep the microresonator thermally stable [8]. Pump laser backward-detuning [29], slow pump modulation [38], and auxiliary laser thermal compensation [39-41] have also been examined to mitigate the thermal destabilization induced by intracavity power decrease. Thermally tuning the resonances through integrated microheaters allows access of the single-soliton microcomb state as well [42]. However, the reported techniques suffer from pump control complexity such as rapid actuators, additional control electronics and lasers, or feedback loops. In addition, the pump-to-microcomb conversion efficiency of the above-mentioned generated soliton microcombs are usually less than 2%, an outstanding challenge due to a limited soliton-pump temporal overlap. Improving phase noise and pulsewidths of the soliton microcomb is also needed for practical application realizations.

Here we report the generation of the self-stabilized soliton microcomb in a strong-coupled dispersion-managed microresonator with a single pump laser, accessing rich and distinct soliton dynamics, sub-100-fs pulse generation, high conversion efficiency, and good $1/f^2$ phase noise performance in a dual-polarized single resonator. First, we report the deterministic dissipative singlet soliton formation and soliton bursts with large soliton existence ranges of ≈ 10 GHz. The remarkable pump-soliton conversion efficiency of 9.2% results from the strongly transverse-magnetic (TM) to transverse-electric (TE) cross-polarization coupling. The ultrafast 77.3-fs pulse generation results from the microresonator dispersion management. Secondly, resonant power switching dynamics are observed for the first time during the pump wavelength forward and backward detuning. The distinct power transients facilitate resonant power buildup and kick out, leading to the formation of dissipative solitons at blue detuning with respect to the resonant transverse-magnetic mode. The pump laser settles in the thermal-locking regime which fundamentally mitigates the characteristic thermal destabilization during soliton formation and results in good phase noise performance. Thirdly, deterministic soliton transitions are observed without requiring stringent control of the pump laser sweeping rate. Two distinct soliton formation routes with the unique soliton cross-phase interaction regime are highlighted which emerge from the chaotically modulated or periodic waveforms. A critical pump power is revealed to separate the formation routes. Fourthly, the coupled supermodes and complex soliton dynamics are examined with an MHz-spectral-resolution optical vector network analyzer. Splitting supermode evolution and a resonance strong-coupling process are experimentally revealed. The cavity resonance, soliton resonance, and breathing signal are also observed. Furthermore, soliton temporal evolution portraits are recorded with a parametric time magnifier to explore the soliton transition dynamics with picosecond temporal resolution, MHz frame rate and sub-ns temporal window.

**Result**

**Self-stabilized soliton microcomb formation in polarization-strong-coupled dispersion-managed nitride microresonators**

A schematic of the self-thermally-stabilized microcomb formation in the dual-polarization strong-coupled silicon nitride microresonator is illustrated in Figure 1a, with a scanning electron micrograph of the microresonator shown in inset **i**. For adiabatic dispersion control, the dual-polarization waveguide width is changed periodically from 1 to 2.5 μm along the microresonator which, for the TE and TM polarizations respectively, features a free spectral range of 19.82 and 19.96 GHz with loaded quality factors of 1.39 and 0.96 million (detailed in Supplementary Materials I). In the microresonator, the TE-polarized anomalous dispersion soliton microcomb has a square hyperbolic secant envelope generated through continuous-wave TM laser pumping in the forward (increasing wavelength) and backward (decreasing wavelength) pump laser detuning. The pump laser is always in the negative-feedback thermally-locked regime – effective pump-resonance blue detuning region at TM polarization – leading to the adiabatic accessibility of the TE-polarized soliton microcomb states at different pump-resonance detunings.

The process of the self-stabilized microcomb formation in the split supermode is illustrated in inset **ii** which includes three physical cavity processes: (1) pump laser power conversion from TM to TE polarization through the TM-TE mode coupling; (2) transient cavity cooling to access the split supermode gap between the $TE_1$ and $TE_2$ during the exit from the $TM_1$ cavity; and (3) negative passive feedback thermal locking to access thermally-stabilized soliton states following the self-thermal locking trajectory in the TM-polarization, keeping at the pump-resonance blue-detuning consistently during backward-pump detuning. The correspondingly measured cold splitting TM supermode spectrum is presented in Figure $1b_1$ along with the converted TE-polarized supermode structure as shown in Figure $1b_2$ via the TE-TM mode coupling. After optically pumping the split supermode, the intracavity power at the TE-polarized mode is recorded by continuously sweeping the pump wavelength along the forward and backward directions as shown in Figure 1c. A remarkable soliton existence range of more than ≈ 10 GHz with stair-like discrete soliton annihilations is observed. The inset is a zoomed-in view of the intracavity power – an extended intracavity field interaction between the TE and TM polarized field – which shows the detailed soliton transition dynamics with a negative slope of the soliton stairs with respect to the pump-resonance detuning, distinct from prior works based on fast laser sweeping [2,6].

We observe that the single-soliton microcomb is formed at the maximum pump-resonance detuning before exiting the TE pump resonance with periodic soliton recoils [23] as shown in Figure 1d, with an expanded view of spectral comb lines shown in the left inset. The soliton microcomb features an enhanced 9.2% conversion efficiency and an ultrafast transform-limited pulsewidth of 77.3 fs. To examine the temporal properties of the soliton microcomb, the intensity autocorrelation is measured by injecting the filtered optical spectrum into a second-harmonic-based intensity autocorrelator. Concurrently, the Turing pattern is soft-excited in the TM polarization with a frequency spacing determined by the TE-TM mode crossing free spectral range. Subsequently, the repetition rate beat note is measured with a high-speed photodetector to examine linewidth and signal-to-noise ratio (SNR) as shown in Figure 1e along with an expanded view of the beat note in the inset. The fitted linewidth is $\approx$ 170 Hz, much narrower than the pump laser linewidth ($\approx$ 100 kHz), indicating excellent coherence from the intrinsic thermal stabilization of the coupled supermodes. The SNR is more than 70 dB over the 100 kHz RF spectral span without additional RF spurs. We also note that the self-stabilized microcomb is operational for long-time periods, preserved in the same state for more than several hours in each measurement study. The repetition rate single-sideband phase noise is measured as shown in Figure 1f. Without locking the pump, the free-running single-sideband phase noise is -108 (-25) dBc/Hz at the 10 kHz (10 Hz) offset frequency of the 19.689 GHz carrier. From 1 kHz to 100 kHz, the measured phase noise falls with a 20-dB/decade slope ($1/f^2$). The self-thermally-stabilized soliton microcomb can be generated with a slow pump laser sweep rate to facilitate the exploration of soliton physics easily.

**Self-stabilized soliton dynamics in the strong-coupled dispersion-managed nitride microresonator**

The soliton mode-locked pulses are spontaneously formed from the extended uncorrelated waveform (chaotic microcomb state) and the periodic waveform (Turing pattern, primary comb line) in the strong-coupled microresonator. During the pump wavelength forward tuning into the TM-polarized supermode, the TM power is gradually converted into the TE-polarization as shown with a positive slope in Figure 2a. A sudden power jump switching behavior is observed with the forward tuning, approaching the parametric oscillation threshold to excite primary comb lines as shown in Figure 2a inset **i**. As the pump wavelength is further tuned, another sudden power jump is observed and the high noise chaotic microcomb is generated as shown in inset **ii**. Then, backward-pump tuning is initiated to decrease intracavity power so that the relative pump-TE-detuning is tuned via differential thermorefractive effects. During the backward-pump tuning, the intracavity power kick out is observed along with the collapse of the extended intracavity waveform into the periodic intracavity waveform. During the backward-pump tuning, the intracavity power gradually decays, clearly distinct from prior works [29]. Inset **iii** of the Figure 2a shows one representative optical spectrum of the mode-locked TE soliton microcomb. Further decreasing the pump-resonance detuning generates the soliton microcomb with characteristic soliton steps as shown in inset **iv**. Concurrently, primary comb lines are formed in the TM polarization.

Figure 2b shows the supermode power transmission without polarization demultiplexing. Along the forward- and backward-pump wavelength tuning, the sudden power injection to the resonant TE supermode and power kick out from the supermode are observed, marked with dashed square boxes. By polarization demultiplexing the TM polarization, the power transmission is recorded as shown in Figure 2c, with the forward- and backward-pump wavelength tuning showing optical hysteresis with negligible TM power dissipation. The characteristic power switching dynamics indicate the exit of the pump laser from the TM resonance and the power kick out from the TM to TE-polarized resonance. Figure 2d shows the optical spectral evolution with respect to pump-resonance detuning, both referenced to the initial backward-tuning wavelength. We observed the characteristic microcomb dynamics from the primary comb lines to the high-noise microcomb state along the forward-pump tuning. In the backward-pump tuning, we further observe the soliton annihilation, collapsing from extended chaotically modulated background to the cross-phase modulated microcombs to the soliton microcomb. The chaotic intracavity waveform provides a statistically random initial environment for soliton microcomb convergence, distinct from the soliton switching dynamics from multiple solitons to the singlet soliton [29]. Correspondingly, the intensity noise power spectral evolution of the microcomb is recorded as shown in Figure 2e. Centered at the pulse repetition rate, a characteristic broadband radiofrequency (RF) spectrum is observed, originating from beating between sub-comb families when the microresonator is operating at the high-noise state. The broadband RF spectrum subsequently collapses into the low-noise repetition tone with the symmetric beat note originating from the degenerate mode interaction between the TE and TM supermodes. Through further tuning of the pump wavelength, a high spectral purity RF signal is observed.

By reducing the pump power, distinct soliton dynamics are observed as shown in Figure 2f where the mode-locked microcomb is directly excited from the organized waveform background (primary comb lines) without accessing the extended chaotically modulated background. The microcomb does not exhibit the sub-comb spectral overlap and enters the low-noise microcomb state directly. The corresponding intensity noise power spectral dynamics are illustrated in Figure 2g where it clearly shows the spectral dynamics without experiencing the high-noise state. This dynamical evolution without the high noise state benefits from the additional thermally-induced mode-crossing phase shift [43]. Figure 2h shows the TE and TM intracavity power evolution with characteristic power jump steps on the switching dynamics. The TM-polarized transmission clearly illustrates the forward- and backward-tuning trajectory and indicates that consistent pump –TM-resonance blue detuning of the soliton microcomb. The TE-polarized transmission shows the observable power switching including the strong-coupling-induced pump-to-microcomb conversion, the power kick out from the high-noise microcomb state, and soliton annihilation. Figure 2i illustrates the TM and TE transmission, and the TE-polarized intracavity power – evidencing the direct transition from the primary comb lines to the mode-locked soliton state. The pump laser exiting the $TM_1$ resonance induces the decrease of intracavity power, directly accessing the soliton state (without the intermediate high-noise TE microcomb state) through the thermorefractive effect. With backward tuning, the power in the TE and TM supermodes gradually decreases to eventually access the soliton states.

**Probing cavity thermal and ultrafast dynamics with an optical vector network analyzer (OVNA)**

The cavity supermode responses are recorded via a phase-modulation-based optical vector network analyzer (OVNA) [44]. Figure 3a shows the experimental setup to record the cavity supermode evolution mediated by the differential thermorefractivity, the cavity power transmission, the TE-polarized intracavity power, the intensity noise spectral and the optical spectral evolution. The OVNA allows us to unveil the cavity supermode evolution with 1.67 MHz RF spectral resolution during the pump forward and backward detunings. The phase modulation is converted into intensity modulation via the asymmetric linear and nonlinear phase responses of the intracavity waveforms or the asymmetric magnitude response of the cavity supermode for two phase-modulated sidebands. Insets **i** and **ii** of Figure 3a show the TE- and TM-polarized split supermode spectra along with the calculated supermode spectra (solid lines) from coupled mode theory. The mode-coupling regimes with different coupling rates are presented in insets **iii** and **iv**. After accessing the strong-coupling regime, the TM power is converted into the TE polarization efficiently. Through the OVNA, Figure 3b shows the microresonator supermode evolution along with the pump forward-backward detuning at five consecutive stages: ① the blue-detuned split cavity supermode (TE$_1$ and TE$_2$); ② the strong TM-TE mode coupling regime when the TE and TM modes are tuned close to degeneracy [45]; ③ the transient thermal cooling to access the red-phase-detuning of TE$_1$ and blue-phase-detuning of TE$_2$; ④ the interaction region between the TE intracavity field and the TM intracavity field; and ⑤ the soliton microcomb in the TE polarization and Turing pattern in the TM polarization.

In regime ②, the TE intracavity power converted from TM polarization is accumulated and approaches parametric oscillation threshold. Subsequently, in regime ③, the microcomb spontaneously jumps into the high-noise state facilitated by supermode blue-shift due to the cavity thermal cooling. The representative intracavity responses are illustrated in Figure 3c. After accessing the increased intracavity power in ②, the split supermode transitions into the singlet soliton mode. The intracavity power is sufficient to support the high-noise microcomb formation. The measured magnitude response in ③ of Figure 3c is broadened due to the broadband nonlinear intracavity fields after accessing the high-noise state. Subsequently, the magnitude response collapses into the narrow bandwidth mode spectrum with a beating frequency signal (breathing frequency) and some additional small beating signals as shown in Figure 3c ④. The magnified inset of Figure 3c ④ shows the cavity resonance, soliton resonance and breathing signal where the cavity and soliton resonances are overlapping. By further backward tuning of the pump wavelength, a decrease in magnitude of S resonance *relative* to C resonance is observed, indicating the soliton annihilation. Figure 3d shows the corresponding TM-polarized supermode spectral evolution investigated using the OVNA. The supermode is accessed by gradually decreasing the pump-resonance detuning which evolves from the weakly-coupled regime ① to the strongly-coupled regime ②. In the strongly-coupled regime, the TM-polarized pump laser is gradually converted into the TE-polarized mode, indicated by the decrease of the power response magnitude. Figure 3e shows the representative power spectral responses at different detunings. In regime ④, the intracavity power response shifts towards the high frequency side correlating with increased pump-resonance detuning. A beat note is also observed with increased magnitude. In regime ⑤, the pump-resonance detuning is continuously increased with the pump exiting the resonance.

**Spatiotemporal coupled Lugiato-Lefever equation modeling**

Supporting the observed dual-polarization supermode soliton dynamics, we examined a modified coupled Lugiato-Lefever equation (LLE) with asymmetric cross-phase modulation coefficients between the TE and TM supermodes (detailed in Materials and Methods). The two LLEs correspond to the TE- and TM-polarized intracavity fields. Since we have only a single pump driving term, additional relative pump-resonance detuning is introduced between the two LLEs. Figure 4a shows the intracavity temporal waveform evolution with respect to the pump-resonance detuning in the TE polarization, with the TM pump. The two distinct cavities ($TE_1$ and $TE_2$) are accessed by sweeping the pump forward and backward. With forward tuning, the high-noise microcomb state is accessed in the $TE_2$ resonance after passing the $TE_1$ resonance where the periodic intracavity pattern is excited. The pump wavelength is subsequently backward tuned to access the soliton state corresponding to the soliton dynamical evolution in Figure 2d. To further explore the pump power dependence of the soliton dynamics, the TE-polarized intracavity temporal waveform evolution is simulated with a lower pump power while fixing all other parameters. Due to the decrease of the pump power, the microcomb states is not excited in the $TE_1$ resonance. Subsequently the high-noise microcomb state is generated in the $TE_2$ resonance. Backward detuning begins at 30,000 roundtrips and subsequently the multiple soliton and single soliton microcombs are obtained as shown in Figure 4b. Figure 4c shows the intracavity power evolution in the TE-polarized supermode, dependent on the pump-resonance detuning at different pump laser powers. By increasing the pump laser power, the nonlinear intracavity field is excited in the $TE_1$ resonance and the high-noise chaotic microcomb existence range is expanded. By increasing the TE-TM resonance coupling ratio $\alpha_{TM\_TE}$, more TM-polarized pump power is converted into TE-polarized power which acts as the TE pump. We also note that, in the measurements, the coupling ratio is continuously changed via the thermorefractive contribution during changing the pump-resonance detuning.

**Real-time observations of soliton transitions with a parametric temporal magnifier**

We developed a parametric temporal magnifier (PTM) [46] based on four-wave mixing in a highly nonlinear fiber to record the slowly evolving soliton transitions in the strongly-coupled microresonator. Compared to dual-comb metrology [47], this has a larger temporal recording window. The 100-ns-long soliton transition portraits, including soliton transition dynamics from a triplet soliton molecule to a doublet soliton and from a doublet soliton molecule to a singlet soliton state, are recorded. Figure 5a shows the measurement schematic wherein the linear frequency chirp is imparted to the truncated soliton waveform via the four-wave mixing process and is subsequently compensated by a dispersive element with opposite dispersion slope. The temporal imaging condition is satisfied with the relation $-1/\varphi_2'' + 1/\varphi_3'' = 1/\varphi_f''$, where $\varphi_2''$, $\varphi_3''$ and $\varphi_f'' = \varphi_1''/2$ are the signal, idler, and pump group-delay dispersion respectively. The temporal magnification ratio of the PTM can be described with $M = \varphi_2''/\varphi_3''$. The PTM output signal is detected by a high-speed photodetector and then digitized and recorded by a high-sampling-rate real-time oscilloscope. We utilized the distinct voltage edge induced by soliton steps to trigger the oscilloscope to precisely capture the soliton transition dynamics. The distinct soliton steps along the pump backward-forward detuning are illustrated in Figures 5b and 5c where the single soliton step is marked with a dashed line box and the soliton breaking behavior from the singlet soliton to the high-noise state is observed. To demonstrate the PTM capabilities, panoramic temporal imaging is conducted with an under-sampling method to record the entire intracavity waveform evolution along the pump backward detuning, initiated from the chaotic microcomb states to the single-soliton state. The intracavity power evolution is recorded simultaneously, overlapping with the temporally magnified waveform denoted with a transparent gray curve.

To clearly visualize the soliton transitions, we recorded the waveform time-trace with a high temporal resolution and a small temporal window. The measured one-dimensional (1D) time-trace data is subsequently converted into a 2D metric data determined by microresonator roundtrip time (50.8 ps). By stitching segmented temporal evolution frames determined by the frame period (40 ns) of the PTM, two panoramic temporal evolution portraits are reconstructed as shown in Figure 5e and 5f. A transition process evolving from a triplet to a doublet soliton molecule is recorded during the swept pump-resonance detuning, in which we observed soliton peak power fluctuations and the absence of soliton temporal drift. The integrated intracavity power over each roundtrip is included with the transparent gray curve showing the soliton decay evolution. Moreover, another soliton transition from a doublet molecule to a singlet soliton state is recorded without soliton fusion. The two solitons exist simultaneously in the microresonator with the temporal separation of ≈ 5 ps and subsequently evolve into the single-soliton with gradual evolution indicated by the negative curve slope of the cross-polarization soliton dynamics.

**Discussion and conclusion.** In this work we demonstrate the generation and real-time observation of the self-thermally-stabilized soliton microcomb in a strongly-coupled dispersion-managed microresonator. The soliton formation in the cross-polarization supermodes utilizes the power conversion process between diverse polarizations, avoiding the thermal instabilities and circumventing the biggest obstacle of soliton formation in the single anomalous-dispersion resonances. The supermode soliton microcomb has an extended soliton existence range, and we have shown that it can emerge from either a stochastic or a periodic background by backward-tuning via an adiabatic process, different from prior works [2,4,5,6,7,8,9,40]. Furthermore, our soliton microcomb has demonstrated one of the highest conversion efficiency at 9.2% [2,4,5,7,8,9,48] and has a tapered and engineered cavity group velocity dispersion is close to zero, enabling sub-100 fs ultrafast pulse formation. With the self-stabilization, the demonstrated soliton microcomb has good free-running phase noise compared to prior works [6,8], and does not require an additional laser [39-41] nor additional nonlinear process for thermal stabilization [49]. The demonstrated cross-polarized deterministic soliton formation can also be applied to generate microcombs wherein the intrinsic cavity thermal noise is the fundamental bound on the repetition-rate phase noise [50] further benefiting low-phase noise millimeter [51,52] and sub-millimeter wave generation [21]. Besides the mitigated thermal dynamics and the real-time observed soliton dynamics, the large soliton existence range can also further decouple the noise transduction processes such as pump-to-cavity repetition rate noise conversion [53] and thermal and nonlinear processes [54]. The demonstrated polarization-diverse soliton microcomb deterministic transitions and dynamics contributes to the fields of chip-scale ultrashort pulse generation, frequency and precision metrology, communications, spectroscopy and information processing.

**Methods**

**Integrated nonlinear dispersion-managed microresonator fabrication:** The fabrication procedure of the microresonator starts with a 5-μm thick $SiO_2$ layer that is first deposited via plasma-enhanced chemical vapor deposition (PECVD) on a *p*-type 8" silicon wafer to serve as the under-cladding oxide and to suppress the substrate loss. An 800-nm silicon nitride layer is deposited via low-pressure chemical vapor deposition (LPCVD) and the resulting nitride is patterned by optimized 248-nm deep-ultraviolet lithography and etched down to the buried oxide cladding via an optimized reactive ion dry etching (RIE). The etched sidewalls have an etch verticality of 88° characterized by transmission electron microscope. Annealing was applied for 3 hours at a temperature of 1150 °C to reduce waveguide propagation loss. Adiabatic mode converters are then implemented to improve the coupling efficiency. The coupling loss from free space to bus waveguides is less than 3 dB per facet. Finally, the nitride microrings are then over-cladded with another 4.5 μm thick oxide layer deposited initially with LPCVD for 0.5-μm and then with PECVD for 4.0-μm.

**Soliton microcomb numerical modeling in the coupled microresonator:** The cavity soliton dynamics are numerically modeled with the spatiotemporal coupled Lugiato-Lefever equation (C-LLE) after taking the anomalous and normal group velocity dispersion for the transverse electrical (TE) and transverse magnetic (TM) modes and the pump laser power conversion into account. The dispersion-managed soliton microcomb dynamics are numerically examined with the equations

$$T_{R\_TM}\frac{\partial}{\partial t}A_{TM}(t,\tau) = -\left[\frac{\alpha_{c\_TM}+\alpha_{p\_TM}}{2} + j(\delta_{TM}-\delta_{TH\_TM}) - j\frac{\beta_{2TM}}{2}L_{cav\_TM}\frac{\partial^2}{\partial \tau^2}\right]A_{TM}(t,\tau)$$
$$+j\frac{n_2\omega L_{cav\_TM}}{c}\left(P_{TM}(t,\tau) + 2g_{TE\_TM}P_{TE}(t,\tau)\right)A_{TM}(t,\tau) + \sqrt{\alpha_c}A_{TM\_P} \quad (1)$$

$$T_{R\_TE}\frac{\partial}{\partial t}A_{TE}(t,\tau) = -\left[\frac{\alpha_{c\_TE}+\alpha_{p\_TE}}{2} + j(\delta_0+\delta_{TE}-\delta_{TH\_TE}) - j\frac{\beta_{2TE}}{2}L_{cav\_TE}\frac{\partial^2}{\partial \tau^2}\right]A_{TE}(t,\tau)$$
$$+j\frac{n_2\omega L_{cav\_TE}}{c}\left(P_{TE}(t,\tau) + 2g_{TM\_TE}P_{TM}(t,\tau)\right)A_{TE}(t,\tau) + \alpha_{TM\_TE}\sqrt{\alpha_c}A_{TM\_P} \quad (2)$$

$$\frac{\partial \delta_{TH\_TM}}{\partial t} = \frac{1}{C_{p\_TM}}\left[\alpha_{T\_TM}\frac{Q_{TM}}{Q_{in\_TM}}\left(P_{TM}(t,\tau)+P_{TE}(t,\tau)\right) - K_{TM}\delta_{TH\_TM}\right] \quad (3)$$

$$\frac{\partial \delta_{TH\_TE}}{\partial t} = \frac{1}{C_{p\_TE}}\left[\alpha_{T\_TE}\frac{Q_{TE}}{Q_{in\_TE}}\left(P_{TE}(t,\tau)+P_{TM}(t,\tau)\right) - K_{TE}\delta_{TH\_TE}\right] \quad (4)$$

where $T_{R\_TM(TE)}$ is round-trip time of the TE or TM cavity, $A(t,\tau)$ is the intracavity waveform electric field amplitude, $t$ is the slow time describing the microcomb dynamical evolution over round trips, $\tau$ is the fast time of the intracavity waveform defined in a reference frame moving at the light group velocity, $\alpha_p$ is the propagation loss, $\alpha_c$ is the coupling loss, $\delta_{TM/TE}$ is the pump-resonance detuning in the TM or TE polarization. $\delta_{TH\_TM/TH\_TE}$ is the pump-resonance thermal detuning in the TM or TE polarization, $\beta_{2TM/2TE}$ describes the second order dispersion coefficient, $L_{cav\_TE/TM}$ is the cavity length, $g_{TE\_TM/TM\_TE}$ is the cross-phase modulation of the electric field at the two polarization directions, $\alpha_{TM\_TE}$ is the power conversion coefficient from the TM to TE polarization which is temperature dependent, $A_{TM\_P}$ is the external pump laser amplitude at the TM polarization, $\delta_0$ is the initial phase offset between the two polarizations, $\alpha_{T\_TM/TE}$ is the temperature coefficient at the two polarizations, $C_{p\_TM/TE}$ is the thermal capacity, $K_{TM/TE}$ is the thermal conductivity, and $Q(Q_{in})$ are the loaded and intrinsic quality factors at the two polarizations. To explore the microcomb formation dynamics in the strong-coupled microresonator, we solve the coupled LLE via a split-step Fourier approach with a time step size of $T_R$ and 512 modes centered at the pump mode. The simulation starts from vacuum noise and runs for $1 \times 10^5$ roundtrips until the simulated solution reaches steady state. $\beta_2$ are -2.27 fs$^2$/mm and 41.07 fs$^2$/mm for the TE and TM polarizations respectively based on our experimental cavity mode characterization which is close to the numerically calculated cavity mode dispersion (detailed in Supplementary Materials I).

**Parametric time magnifier for real-time dynamics measurement:** A parametric time magnifier (PTM) is developed to record the 2D ultrafast dynamical soliton transitions in the coupled dissipative soliton microcomb. The PTM is implemented though parametric four-wave mixing (FWM) in a highly nonlinear fiber (OFS) with a length of 50-m which has a zero-dispersion wavelength of 1556 nm, a dispersion slope of 0.019 ps/(nm$^2$·km) and nonlinear coefficient of 11.5 W$^{-1}$km$^{-1}$. A 250 MHz stabilized femtosecond fiber laser (Menlo FC1500-250-WG) is utilized as the FWM pump after electro-optic pulse picking and spectral filtering. An arbitrary waveform generator (Tektronix AWG520) synchronized by the pulse fiber laser is used to generate a square pulse pattern to realize the pulse picking reducing the repetition rate to 25 MHz. A wavelength-division multiplexer (WDM) is used to filter out the optical spectrum from 1554 to 1563 nm which is subsequently amplified by a C-Band amplifier to around 100 mW. The excess amplified spontaneous emission is suppressed by another WDM with identical spectral characteristics. Additionally, the soliton microcomb is intensity modulated with the same pulse picking frequency at 25 MHz and spectrally filtered from 1543 to 1547 nm to be the FWM signal. Before combining the FWM pump and signal, they are chirped by two spools of dispersion compensated fiber with dispersion of -52.5 ns/nm/km and -26 ns/nm/km, respectively. The generated idler is filtered by a bandpass filter and is subsequently amplified by a L-band optical amplifier. A spool of dispersion compensating fiber with -2040 ns/nm/km dispersion is used to counteract the chirp of the filtered and amplified idler signal. An 18 GHz photodiode (ET-3500F) is used to detect the time magnified signal and a 100 GS s$^{-1}$ oscilloscope (Tektronix MSO 72004 C) is used for digitization. The PTM is of 1.5 picosecond temporal resolution and 500 ps FWHM temporal field-of-view. The measured temporal magnification of the PTM is $M = 73.2$.

**Acknowledgements:** We acknowledge financial support from the Office of Naval Research (N00014-16-1-2094), Lawrence Livermore National Laboratory (B622827), and the National Science Foundation (1824568, 1810506, 1741707, 1829071).

**Author contributions:** W.W. and X. J. conducted the experiments. W.W and H. Z analyzed the data and performed the simulations. A.K.V. contributed to the experiments. M.Y., G.-Q.L. and D.-

L.K. performed the device nanofabrication. W.W., and C.W.W. initiated the project. W.W., T. M. and C.W.W. wrote the manuscript. All authors discussed the results.

**Competing interests:** The authors declare that they have no competing interests.

**Data and materials availability:** All data needed to evaluate the conclusions in the paper are present in the paper and/or the Supplementary Materials. Additional data related to this paper may be requested from the authors.

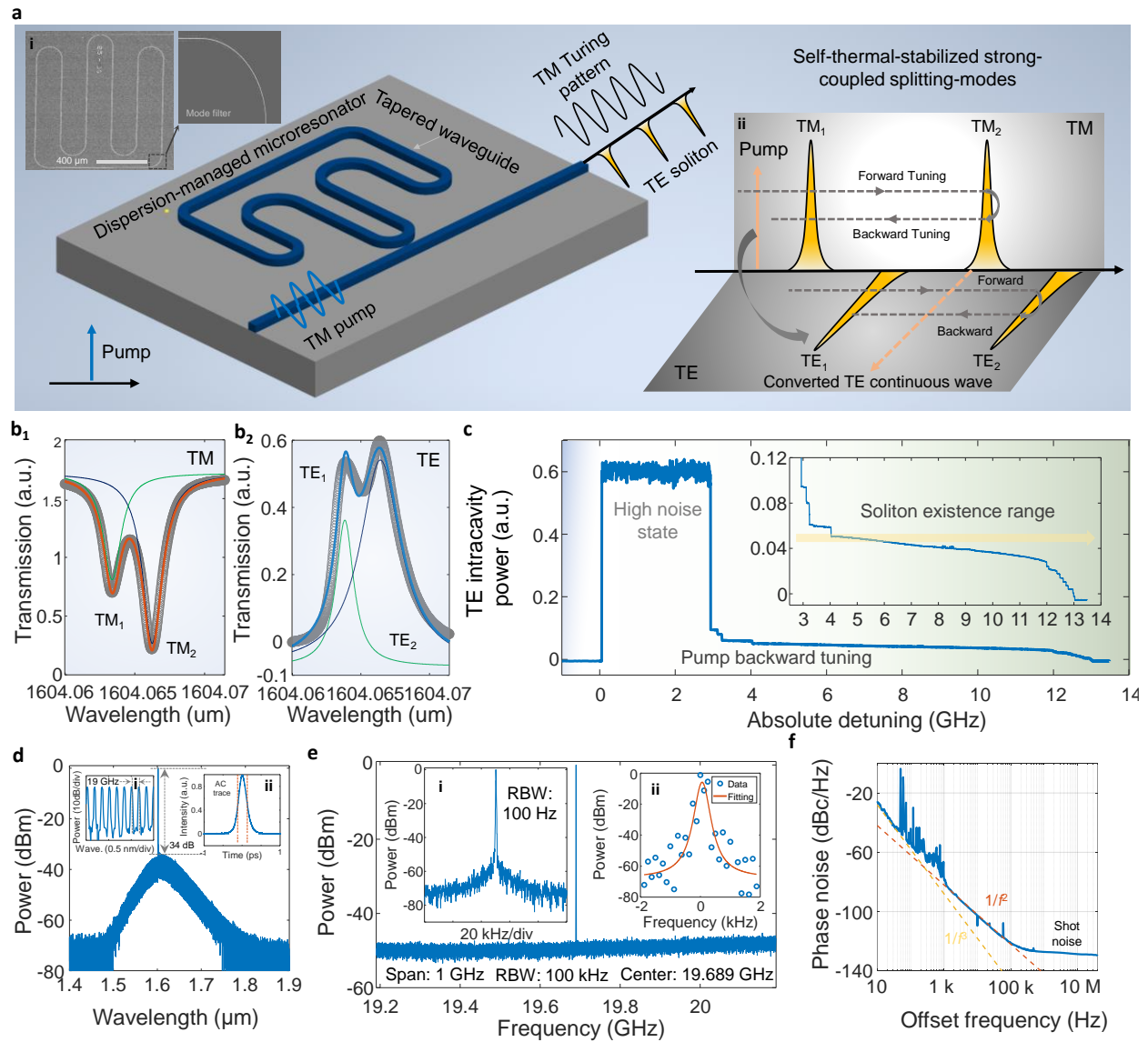

**Figure 1 | Polarization-diverse self-stabilized soliton formation in the strongly-coupled dispersion-managed nitride microresonator. a**, Schematic of the self-thermally-stabilized microcomb generation where a transverse magnetic (TM) polarized continuous wave laser drives a dispersion-managed microresonator leading to the generation of the transverse electric (TE) polarized soliton microcomb and TM-polarized Turing pattern. Inset **i**: Scanning electron micrograph of the dispersion-managed microresonator including expanded view of the designed mode filters. Inset **ii**: Conceptual illustration of the strongly-coupled splitting modes for formation of the self-thermally-stabilized microcomb. **b$_1$** and **b$_2$**, TM-polarized cold resonant supermode, and the converted TE-polarized cold resonant supermode. **c,** Intracavity power in the TE polarization with respect to the backward-pump wavelength detuning. Inset: Soliton existence range indicating the transition dynamics. **d,** Generated self-thermally-stabilized soliton microcomb with low repetition rate, an enhanced pump-microcomb conversion efficiency of 9.2% and a transform-limited full-width half-maximum pulse width of 77.3 fs. Inset **i**: the zoomed optical spectrum of the soliton microcomb with a free-spectral range of ≈ 19 GHz. Inset **ii**: the measured intensity autocorrelation trace. **e,** Measured repetition rate tone. Inset **i**: the zoomed repetition rate beat note. Inset **ii**: the fitted linewidth showing the 3-dB linewidth of 169 Hz over 1 ms observation time. **f,** Measured single-sideband phase noise of the free-running microcomb after optical-to-electrical conversion.

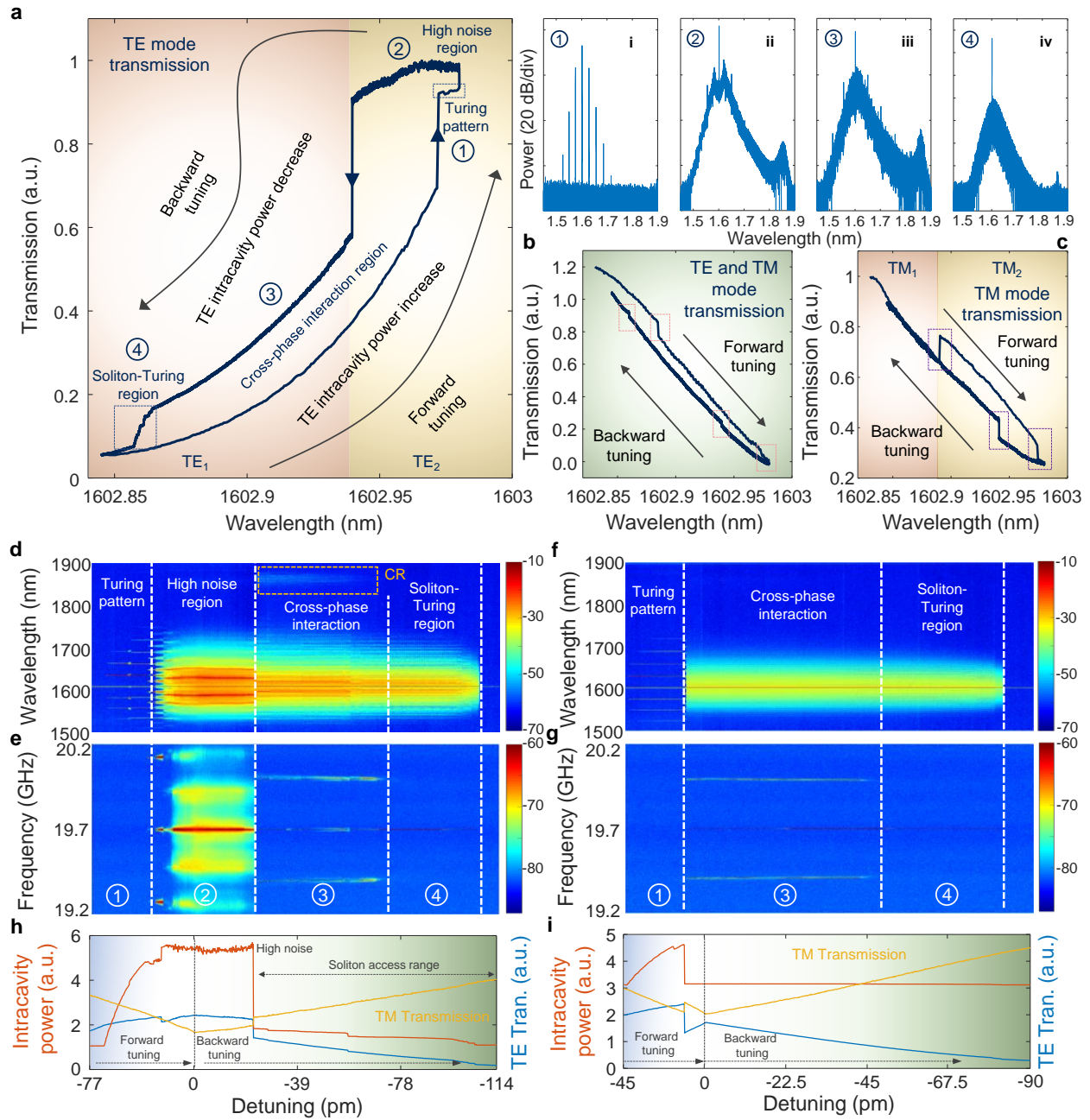

**Figure 2 | Self-stabilized cross-polarized soliton microcomb dynamics in the strongly-coupled dispersion-managed microresonator. a,** TE-polarized cavity power transmission along the bidirectional TM-polarized pump wavelength detuning. The intracavity power is built up via the polarization conversion process leading to the strongly coupled transient and two-step intracavity power enhancement during the pump-wavelength forward detuning. The cross-phase interaction between the TE-TM mode families after intracavity power 'kicking' out and the soliton

annihilation regime with the characteristic staircase pattern during the pump-wavelength backward detuning are observed. The measured optical spectra at the different microcomb states represented in **a** as shown in insets **i-iv**: the primary Turing pattern microcomb, the high-noise chaotic microcomb, the TE-TM coupled low-noise microcomb (soliton mode-locking state, TE-TM dual microcomb), and the thermally-stabilized TE soliton microcomb. **b,** The power transmission without polarization demultiplexing during the pump-wavelength roundtrip detuning shows the switching dynamics between the TE and TM supermodes. The characteristic transient power jumps are highlighted in the dashed orange boxes. **c,** The power transmission in the TM polarization only where the switching dynamics between the TE and TM supermodes are marked with dashed purple boxes. **d,** The optical spectral evolution with respect to pump-resonance detuning, referenced to the initial backward tuning wavelength, illustrating the soliton microcomb dynamics. In the backward-pump laser detuning, the cross-polarized microcomb emerges from the extended chaotically modulated background. **e,** The intensity noise power spectral evolution corresponding to **d,** showing the broadband noise spectrum resulting from the beating of subcomb lines ②, repetition rate linewidth broadening and extra beat notes ③, and high spectral purity repetitive rate signal ④. **f,** The optical spectral evolution where the soliton microcomb emerges from the periodic background without excitation of the chaotic microcomb state. **g,** The intensity noise power spectral evolution corresponding to **f** denoting low-noise cavity dynamical evolution. **h,** The color-coded TE and TM power transmission and the TE-polarized intracavity power versus pump-resonance detuning corresponding to **d** and **e**, indicating the large soliton existence range of more than 10 GHz and the power conversion processes. **i,** The respective color-coded TE and TM power transmission and the TE-polarized intracavity power without the high-noise state corresponding to **f** and **g**.

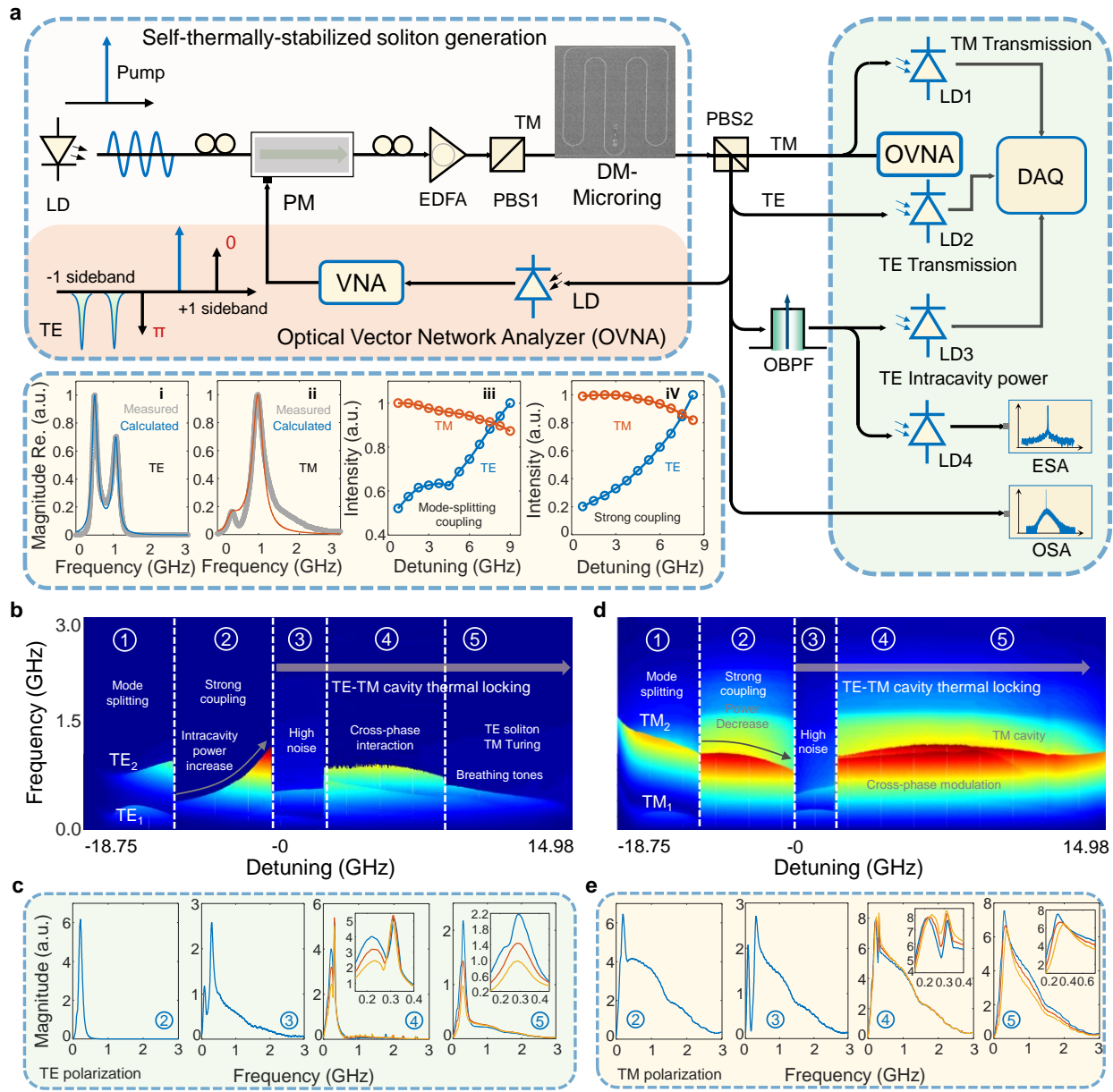

**Figure 3 | Probing the supermode cavity thermal and ultrafast dynamics with an optical vector network analyzer (OVNA). a,** The experimental setup of the self-stabilized microcomb characterization. Inset **i-ii**: The microresonator supermodes in the TE and TM polarization showing the clear split mode spectra. Inset **iii-iv**: The normalized power conversion processes in the weak and strong coupling regimes corresponding to ① and ② at **b** and **d**. **b,** The dynamical cavity supermode evolution with respect to the pump-resonance detuning in the TE polarization measured via the OVNA including ①: the evolved split mode spectra region, ②: the polarization strongly-

coupled region, ③: the high noise region, ④: the dual-mode interaction region, ⑤: the solitary wave region. **c,** The representative cavity mode spectra in **b** including the strongly converted TE mode ②, the high noise mode spectrum including the cavity mode spectrum and the nonlinear broadening mode spectrum induced by stochastic phase response ③, the soliton-induced nonlinear mode spectrum (*S*-resonance) and the breathing signal ④, ⑤. The *S*-resonance magnitude depends on the soliton number which overlaps with the cavity mode spectrum (*C*-resonance). Insets are the expanded views. **d,** The dynamical cavity supermode evolution in the TM polarization showing the power decay region ②, the high noise region ③, the dual-mode interaction region ④ and ⑤. **e,** The representative cavity mode spectra, and the nonlinear mode spectra induced by the phase response of the intracavity field at the different regions.

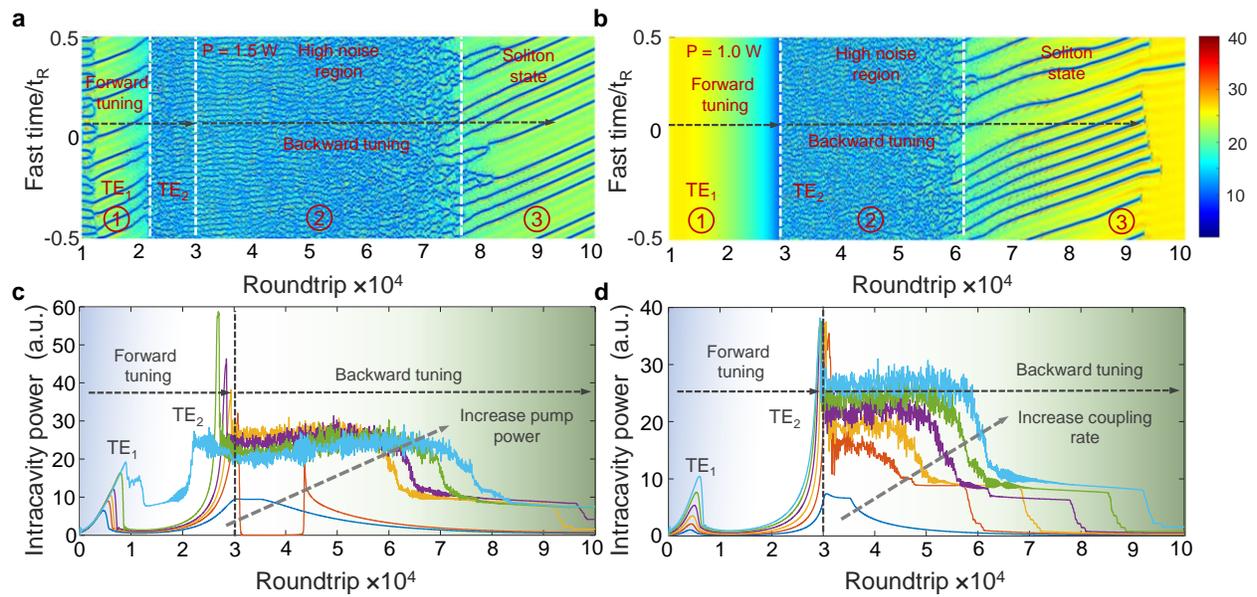

**Figure 4 | Spatiotemporal LLE modeling of the self-stabilized cross-polarization supermodes in both forward- and backward-detuning. a,** The intracavity waveform evolution along the pump wavelength bidirectional detuning at the cavity supermodes (TE$_1$ and TE$_2$) when pump power is set to 1.5 W. **b,** The intracavity waveform evolution when the pump power is set to 1.0 W. **c,** The intracavity power of the TE-polarized waveform evolution when sweeping pump power from 0.7 to 1.5 W. **d,** The intracavity power of the TE-polarized waveform evolution when sweeping the coupling rate between the TE- and TM-polarized supermodes from 0.0032 to 0.0097.

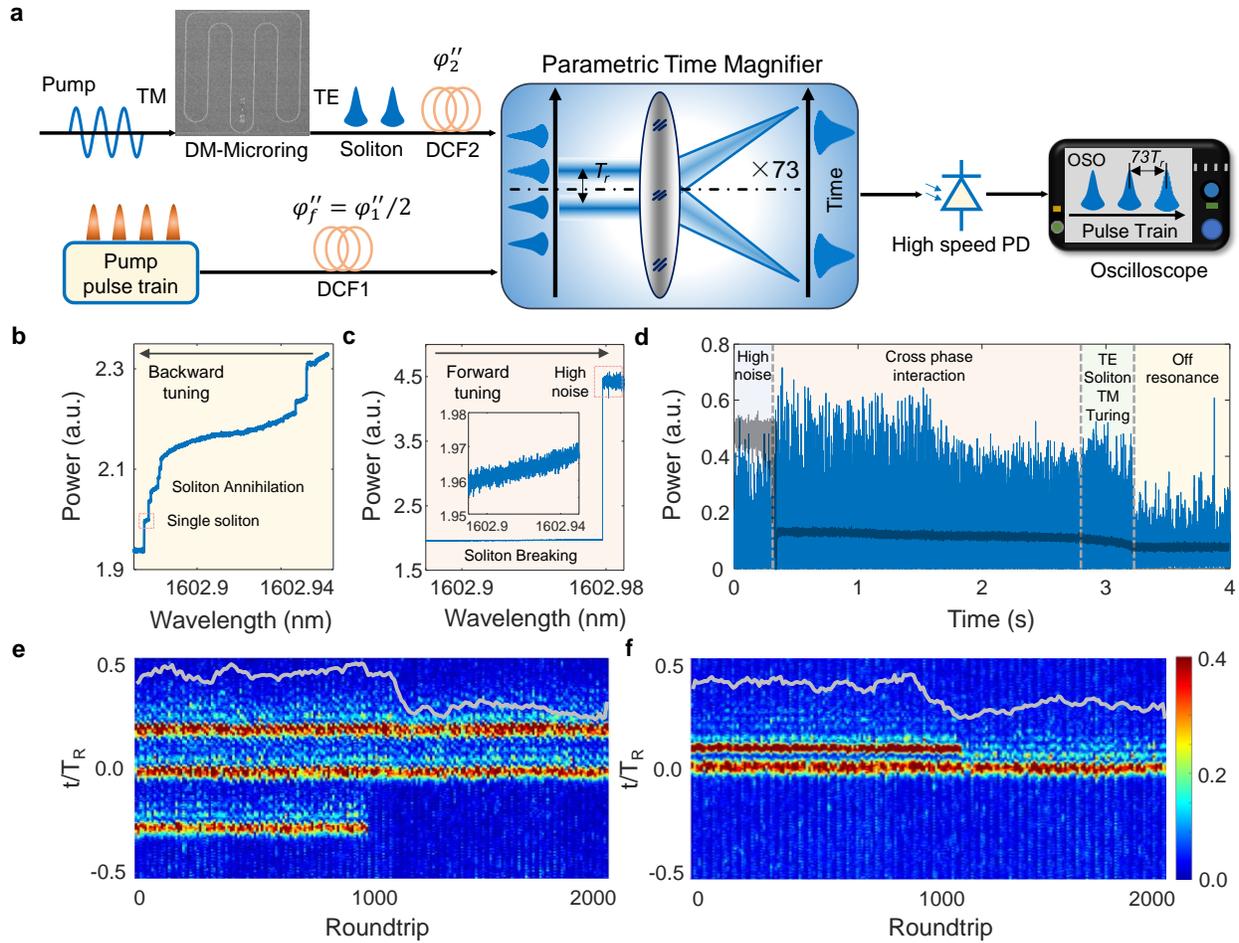

**Figure 5 | Real-time observations of the self-stabilized cross-polarization soliton molecule transitions with a picosecond parametric temporal magnifier.** The fast temporal axis is the integrated intracavity power over each roundtrip; the slow temporal axis is the cavity roundtrip evolution. **a,** The experimental setup for recording soliton transitions including a stabilized femtosecond fiber laser, dispersion compensated fiber (DCF), Kerr soliton train, four-wave-mixing-based parametric time magnifier, a high-speed photodetector, and a high-speed oscilloscope. **b,** The intracavity power transition along the backward-pump laser detuning showing the soliton annihilation evolving from the high-noise chaotic state to the cross-phase-modulation interaction state to the dissipative Kerr soliton state with characteristic soliton steps. **c,** The intracavity power transition along the forward-pump laser detuning after accessing the singlet soliton state where the soliton breaks from the singlet state into the high-noise chaotic state. **d,** The under-sampled intracavity dynamics during the backward-pump laser detuning evolves from the high-noise state to the cross-phase-modulation interaction state to the single-soliton microcomb
27

state. **e** and **f**, The transitory observations of the triplet-to-doublet soliton molecule states and the doublet-to-singlet soliton states recorded by a parametric time magnifier with a 25 MHz frame rate. The 100-ns transition portraits are observed by stitching multiple sub-ns temporal windows. Shown in the vertical axis is the fast temporal axis – the measured integrated intracavity microcomb power over each roundtrip – versus the evolved roundtrips as the horizontal slow temporal axis.